\documentclass[11pt]{article}
\usepackage{appb,epsfig}
\begin{document}
%
\title{TRANSPORT PROPERTIES
AND NEUTRINO EMISSIVITY
OF DENSE NEUTRON--STAR MATTER
       WITH LOCALIZED PROTONS}
\author{D.A. BAIKO
\address{
        A.F. Ioffe Physical Technical Institute,
        194021, Politekhnicheskaya 26, St.Petersburg, Russia \\
        N. Copernicus Astronomical Center,
        Polish Academy of Sciences, Bartycka 18,
        00-716 Warszawa, Poland \\
        e-mail: {\tt baiko@astro.ioffe.rssi.ru}}
\and P. HAENSEL
\address{
        N. Copernicus Astronomical Center,
        Polish Academy of Sciences, Bartycka 18,
        00-716 Warszawa, Poland \\
        e-mail: {\tt haensel@camk.edu.pl}}
}
\headtitle{Transport properties of dense neutron-star
           matter with localized protons}
\headauthor{D.A. Baiko, P. Haensel}
\maketitle

\def\la{\;\raise0.3ex\hbox{$<$\kern-0.75em\raise-1.1ex\hbox{$\sim$}}\;}
\def\ga{\;\raise0.3ex\hbox{$>$\kern-0.75em\raise-1.1ex\hbox{$\sim$}}\;}

\begin{abstract}
As pointed out by Kutschera and W{\'o}jcik,
very low concentration of protons
combined with a specific density dependence  of effective neutron--proton
interaction could lead to a localization of ``proton impurities''
in neutron medium at densities exceeding
four times normal nuclear matter density.
We study consequences of the localization of protons
for transport processes in dense neutron star cores, assuming
random distribution of proton impurities. Kinetic equations,
relevant for the transport of charge, heat and momentum, are solved
using variational method. Localization of protons
removes a $T^{-2}$ factor from the transport coefficients,
which leads, at lower temperatures,
to a strong decrease of thermal conductivity, electrical conductivity
and shear viscosity of
neutron star matter, as compared to the standard case, where
protons form a Fermi liquid.
Due to the localization of protons a number of conventional
neutrino emission processes
(including modified URCA process) become inoperative
in neutron star cores. On the other hand,
the energy loss rate from neutrino--antineutrino pair
bremsstrahlung due to
electron and neutron scattering off (localized) protons,
will have a specific $T^6$ dependence,
which could modify the cooling
of the neutron star core, as compared to the standard case.
Possible astrophysical implications of the localization of protons
for neutron star evolution and dynamics are discussed.
\end{abstract}

\section{Introduction}
The composition of neutron star matter above
three times normal nuclear density ($\rho > 3\rho_0$,
where $\rho_0=2.8\times 10^{14}~{\rm g~cm^{-3}}$)
is largely unknown.
Some many--body calculations suggest, that above
$3 \rho_0$
the baryon component of
matter consists nearly exclusively of
neutrons, with a small admixture (a few percent) of
protons (see, e.g.\ models UV14+TNI, AV14+UVII of \cite{WFF88});
such a composition would be similar to that at
$\rho \simeq \rho_0$. A small admixture of protons in high--density
neutron matter could behave quite differently than at densities
$\rho \simeq \rho_0$. As shown by Kutschera and W{\'o}jcik
\cite{KW93}, coupling of ``proton impurities'' to the density waves in
neutron matter could lead, above some critical density,
to localization of protons in the potential
wells associated with the neutron density inhomogeneities;
at smaller densities the coupling results in a gradual
increase of the proton effective mass \cite{KW93}.
We will show in the present paper, that
the localization of protons would change dramatically
transport properties of neutron star matter.
It would also lead to the ``switching on'' of two new neutrino
emission processes, which would influence
cooling of a neutron star.

In order to visualize possible effect of
the localization of protons, let us consider the standard case,
when nucleons form degenerate normal
Fermi liquids. The transport coefficients of
neutron star matter exhibit then characteristic temperature dependences,
which result from the combined effect of the Pauli principle,
and the energy and
momentum conservation in the scattering processes \cite{LP81}.
The leading terms in the low--temperature expansions of
electrical conductivity, $\sigma$, and shear viscosity, $\eta$, are
proportional to $T^{-2}$, while the low--temperature behaviour
of thermal conductivity, $\kappa$, is given by
$\kappa \propto T^{-1}$.

The conventional neutrino emission processes in the standard
$npe$ matter include the modified URCA process
and the neutrino--antineutrino pair
bremsstrahlung from $nn$, $np$, and $pp$ collisions.
In the absence of nucleon superfluidity
the energy emission rates for all these processes
vary with temperature as $T^8$ (e.g.\ \cite{FM79}).

Localized protons will act as scattering centers for
neutrons and electrons. In what follows, we assume that the
localized protons do not exhibit a long--range
crystalline order  (this point is discussed in
Section 2). In such a case, an elementary consideration indicates,
that the localization would give
$\sigma^{\rm loc.p.},~ \eta^{\rm loc.p.} \propto T^0$, and
$\kappa^{\rm loc.p.} \propto T$; one
may thus expect a strong effect of the proton localization on
the transport coefficients at lower temperatures,
where scattering off protons dominates. As
will be shown in this paper, the localization of protons
produces a drastic decrease of the
transport coefficients of neutron star cores,
as compared to the standard case, when protons form a Fermi
liquid: $\sigma^{\rm loc.p.}/ \sigma,
~ \eta^{\rm loc.p.}/\eta~\sim 10^{-5} - 10^{-6}~T_8^2$, and
~ $\kappa^{\rm loc.p.}/\kappa~\sim 10^{-4}~T_8^2$,
for densities around $4 \rho_0$ and for a proton fraction
about 1 \% ($T_8 \equiv T/10^8~$K).

Furthermore, the emission
of neutrino--antineutrino pairs in the $nn$ collisions
is the only process of the above mentioned, that survives
the localization of protons. Instead, there appear
the neutrino--antineutrino  pair emission from the scattering of
neutrons and electrons off localized protons.
In both cases the rate of this process
reproduces the temperature dependence
of the direct URCA process $(T^6)$, which, if operative,
accelerates drastically the cooling of a neutron star.
As will be seen in Section 6, when neutrons are not
superfluid the neutrino emissivity due to $np$ collisions
is  approximately 3.5 orders of magnitude larger than that due to $ep$
collisions. Despite the same temperature dependence,
the neutrino emission due to neutron scattering off
localized protons is much less efficient than the direct
URCA process.  However, the ratio of the emissivity
due to the $np$ bremsstrahlung to the emissivity
due to the modified URCA process (the most important one among
the standard processes in neutron star cores) could be quite large in the
temperature range of interest:
$Q_{\rm Brem}^{n-{\rm loc.p.}} / Q_{\rm mURCA} \sim 2 \cdot 10^3~T^{-2}_8$,
for the same density and proton fraction.
This implies that the proton localization could lead
to an intermediate regime of a neutron star cooling:
more rapid than the standard cooling, provided by
modified URCA process, and less fast than the accelerated
cooling due to direct URCA process.

The paper is organized as follows.
The physical conditions in neutron star matter with localized
protons are discussed in detail in Section 2.
In particular, we emphasize the similarity between
the behaviour of a proton in neutron matter and a polaron
behaviour of slow electrons in solids;
we also present some arguments against a crystalline ordering
of localized
protons, and discuss the importance of relativistic effects in
the neutron component of matter. Kinetic equations relevant for
the transport of charge, heat and momentum, as well as
variational solutions to them,
and analytical
expressions for the transport coefficients, are derived in Section 3.
Angular averages of
scattering probabilities, appearing in the expressions for
the transport
coefficients, are calculated in Section 4. In Section 5, we
improve the variational solutions to make them asymptotically exact
in the high--temperature regime
and present our results for the transport coefficients in the
form suitable for practical applications.
The neutrino energy emission rates from the $ep$ and $np$
bremsstrahlung are estimated in Section 6.
Finally,
in Section 7 we discuss some astrophysical implications
of these results.

Throughout the paper we will mostly use the units
$\hbar=c=k_{\rm B}=1$
and turn to the normal units
whenever presenting the final results.

\section{Neutron star matter with localized protons}
Consider $npe$ matter at super nuclear densities.
The typical kinetic energy of a proton can be
estimated from the uncertainty principle and reads
\begin{equation}
        T_p \sim {\hbar^2 n_p^{2/3} \over 2 m^\ast_p} =
        0.7 \left({n \over 4 n_0} \cdot {x_p \over 0.01} \right)^{2/3}
        \left(m_p \over m_p^\ast \right) ~~ {\rm MeV},
\label{Tp}
\end{equation}
where $n$ is a mean nucleon density, $x_p$ is a proton fraction
(equal to an electron fraction $x_e$), $n_0$ is the normal nuclear
number density, $n_0 = 0.16~{\rm fm^{-3}}$, and $m_p^\ast$
is the proton effective mass resulting from the two--body
nucleon--nucleon (NN) interactions.

A proton in neutron matter also has an effective potential
energy $V_{\rm eff}$, the value of which depends on the neutron density
and ranges, according to different
parametrization, from 55 to 75 MeV for $n_n \sim 4 n_0$
(fig.\ 1 of \cite{KW93}). Taking into account the possible
inhomogeneity of the neutron sea we can write the Hamiltonian
of a proton in the form
\begin{equation}
    H_p = {-\hbar^2 \nabla^2 \over 2 m_p^\ast} + V_{\rm eff}(n_n) +
          n_n {\partial V_{\rm eff} \over \partial n_n}
          \left({ \delta n_n \over n_n } \right),
\label{Hp}
\end{equation}
where the last term describes the coupling of the proton
to the density waves in neutron matter.
The quantity $\sigma(n_n) = n_n \partial V_{\rm eff} / \partial n_n$
plays role of the coupling strength and is of the order
550 -- 600 MeV at $n_n \sim 4 n_0$. On comparison with
the estimate of the kinetic energy, Eq.\ (\ref{Tp}),
it becomes clear, that the coupling to neutron density waves
might not be neglected, if one aims at a realistic
description of a proton behaviour in high--density
neutron star matter with low proton fraction \cite{KW93}.

The calculations presented in \cite{KW93} indicate, that
at weak and intermediate couplings the proton, interacting
with neutron density waves (which will be referred to as
phonons, while, basically, these are acoustical modes of neutron
matter), acquires an additional
effective mass, which increases gradually with increasing
neutron density.
The situation reminds the so--called ``large'' polaron
in solids, where a slow electron, moving through
a crystal, is dressed into a cloud of virtual phonons,
and, consequently, has an effective mass exceeding its
``bare'' band effective mass. If the coupling strength
increases further, the polarization of the ion
lattice by the electron, could get very strong.
It can eventually be trapped by a local
deformation of the lattice, induced by the electron itself \cite{T61}.
This latter situation corresponds to a ``small'' polaron.
If the temperature
is sufficiently high the electron can be then kicked out
of the potential well trapping it. However, in the low
temperature regime, the electron could only
tunnel slowly through the lattice and spends most of the time
near one ion.

Let us now come back to the proton moving through the neutron
background. The background polarization, induced
by the proton, could be characterized by its spatial scale
$R_p$. With increasing neutron density the proton--phonon
coupling strength $\sigma(n_n)$ increases, whereas
the $R_p$ decreases. The conditions for proton self--trapping
are roughly given by two inequalities:
$m^\ast_p \sigma(n_n) |\delta n_n| R_p^2 / n_n \hbar^2 > 1$,
which ensures, that the potential well formed contains
a bound state for a proton, and $R_p<n_p^{-1/3}$, which
allows each proton to produce its own potential well,
and ensures no overlap between wave functions of different protons.

In order to check if this ``small''
polaron regime occurs for a proton in neutron matter
one have to perform a detailed calculation.
In particular, one should compare the energy of the state with
trapped proton and that of the state,
in which the proton is not localized.
The results of calculations of this type has been
reported in ref.\ \cite{KW93}. These
authors evaluated the energy of a Wigner--Seitz cell
with homogeneous distributions of neutrons and a proton
versus that of the cell with proton wave function
localized near the center of the cell and a
neutron distribution having minimum at the center.
The parameters, characterizing the deviation of those
distributions from the homogeneous ones, has been treated
as variational. It was found, that above some
critical density (ranging from 4 to 9 $n_0$ depending on
the model; from now on we will adopt the most optimistic
model and assume the critical density to be $\simeq 4 n_0$)
there existed a domain of the parameters,
where the localized state was energetically preferable
to the uniform one. Therefore, above this density,
and provided $R_p$, the rms radius of the proton probability
distribution corresponding to the minimum energy,
is sufficiently small, the protons most likely
are in effective potential wells trapping their
wave functions. The typical value of $R_p$
at $4 n_0$ is about 1 fm and typical depth of the well
is $\sim 100$ MeV. The mean distance between protons
\begin{equation}
         a_p \simeq n_p^{-{1/3}} =
                    5.4~\left( {n \over 4n_0}
                    \cdot{x_p \over 0.01}\right)^{-{1/3}}~{\rm fm}~,
\label{d_p}
\end{equation}
while the energy of the zero--point vibration of a proton
in the well is $T_0 \sim 42 (R_p/1 ~{\rm fm})^2$
$(m_p/m_p^\ast)$ MeV.
Thus, we see that at densities
above $4 n_0$ and for sufficiently low proton fraction,
say 1 -- 5 \%, the wave functions of the protons are well
localized around the neutron density minima, and there is
no overlap between them. This, in turn,
means that there is no need to invoke Fermi statistics
to describe the proton system. The temperature, when falls below,
say 10 MeV, is negligible compared to the relevant
energy scale in the proton--trapping well,
so that the protons occupy the lowest available
energy levels, and other (degenerate) particles
scatter off them elastically. The proton could then
only tunnel through the neutron background with very low
probability, and in what follows, we will neglect
the possibility of such events.

Let us discuss in some detail the astrophysical scenario of the
formation of a dense neutron star core, where one might
expect the localization of protons. According
to the standard scenario of the formation of neutron stars in
a gravitational collapse of massive stellar cores, initial
temperature in the central core of a newly born neutron star is
$T \simeq 30~$MeV. Moreover, due to neutrino trapping
in dense hot plasma, initial proton fraction is very high,
$x_{p} \simeq 0.3$. Both a high value of $x_p$
and a high value of $T$ exclude possibility of
the localization of protons. It may take place after
the neutron star core becomes transparent to
neutrinos, and the core temperature falls to the range of
$\la 10$ MeV. One could ask a question, whether the
proton impurities could be ordered during the
localization process due to some long--range interaction?
We immediately find, that Coulomb energy, estimated as,
\begin{equation}
   E_{{\rm C}pp} \sim {e^2\over a_p} \simeq 0.3 \left( {n\over 4n_0}\cdot
                     {x_p\over 0.01}\right)^{1/3}~{\rm MeV}~,
\label{Ecoul}
\end{equation}
is negligible compared to
the potential energy, felt by a localized proton in its effective
well, or the
energy, resulting from the Fermi motion of
protons before the localization point.
However, it is not excluded, that some correlation of proton sites
could be induced by a long--range
correlation resulting from (strong)
nuclear interactions in the neutron sea.
A detailed discussion of this possibility
is difficult and goes beyond the scope of the present paper.
One could recall only, that the localization occurs at rather
high temperature, implying large thermal fluctuations, and it is
uneasy to imagine a long--range correlation,
originating from
strong (that is short--range) forces, with energy scale of several
MeVs. Therefore, we think that neutron star
matter with localized protons is formed as a
disordered system and remains locked in this disordered state
during subsequent cooling. In what follows, we will assume that
the positions of localized protons show no long--range
crystalline order: localized protons will be treated as randomly
distributed, spatially fixed
scattering centers for neutrons and electrons.

Transport processes in neutron star matter with localized protons
are carried out by electrons and neutrons.
For simplicity, we will not allow for the presence of muons.
Their inclusion would lead to some modifications of the kinetic
equation formalism (see, for details, the paper by Gnedin and
Yakovlev \cite{GY95}, where the thermal conductivity of $pe\mu$
component of neutron star core matter was studied under standard
assumptions) but would not alter
qualitatively the main results of this paper.

The Fermi energy of electrons is given by
\begin{equation}
         \varepsilon_{Fe} = 113.4
         \left( {n\over 4 n_0} \cdot {x_p\over 0.01}\right)^{1/3}~
         {\rm MeV}~.
\label{eFermi}
\end{equation}
Typical energy of Coulomb $ee$ and $ep$ interactions
are on the order of $E_{{\rm C} pp}$ (\ref{Ecoul}), that is much
smaller than $\varepsilon_{Fe}$.
Therefore, electrons should
be treated as a free, uniform, ultrarelativistic, and strongly degenerate
($\varepsilon_{Fe}\gg T$)
Fermi gas. Transport of energy, momentum and charge is carried out
by the elementary
excitations in electron gas --- thermal electron quasiparticles. Electron
quasiparticles are approximated  by the excited single--particle
states in a free electron gas, close to the
Fermi surface [i.e. with momenta close to
$k_{Fe} = \left(3\pi^2 n_e\right)^{1/3}$], with
effective mass given by $m^*_e =\varepsilon_{Fe}$.
The pair interaction between electron quasiparticles is described
in the formalism of the dielectric function, the details of which
are outlined in Section 3. Here we mention only, that since usually
we have $T \ll 0.1 \varepsilon_{Fe}$ in the neutron stars cores,
the effect of dynamical screening can be neglected.

Neutrons, by contrast, form a strongly interacting Fermi system.
We will
consider the case, when neutrons are normal (non--superfluid).
Transport processes involving neutrons can then be described in
the spirit of the Landau theory of normal Fermi liquids.
Namely, transport of energy and momentum  is carried out by the
neutron quasiparticles --- elementary single--particle excitations
in the vicinity of the Fermi surface. As we restrict ourselves
to strongly degenerate neutron matter, the gas of neutron
quasiparticles is dilute. The Fermi momentum of neutron
quasiparticles coincides with that of the real neutron matter,
$k_{Fn}= \left(3\pi^2 n_n\right)^{1/3}$. The
neutron quasiparticle velocity at the Fermi surface is given by
\begin{equation}
    v_{Fn} = {\hbar k_{Fn} \over m^*_n}=
    0.56 \left[{n\over 4 n_0}(1-x_p)\right]^{1/3}
    {m_n\over m^*_n}~,
\label{v_Fn}
\end{equation}
where $m^*_n$ is the neutron quasiparticle effective mass.
We see, that at densities of interest
(we will confine ourselves to densities below $5 n_0$),
and with $x_p \ll 1$, neutrons
gradually become relativistic, although not very much;
nevertheless, we will take neutron relativism into account.
In the non--relativistic system the effective mass of neutron
quasiparticles $m^*_n =p_{Fn}/v_{Fn}$ differs
from bare nucleon mass $m_N$ due to many--body
effects. Here it should also include relativistic  effect.
Using Lorentz invariance arguments one
obtains, in the reference frame of neutron matter,
$m^*_n / \varepsilon_{Fn}=1+F^s_1/3$, where
$\varepsilon_{Fn}$ is the neutron chemical potential,
which includes neutron rest mass, and $F^s_1$ is the Landau
parameter.

Actually, the
presence of proton impurities leads to the appearance
of neutron density inhomogeneities.
These inhomogeneities will slightly modify the mean--field,
felt by an incident neutron quasiparticle (the de Broglie wavelength
of which is generally smaller than the sizes of the
inhomogeneities), inducing a continuous
drift of the quasiparticle in the momentum space.
We note, that these
inhomogeneities are restricted
to a very small fraction of the volume $(R_p/a_p)^3\sim 0.01$,
and are concentrated around the proton impurities.
The latter produce sudden changes in neutron momenta,
thus making a dominant contribution to the scattering.
The contribution from the inhomogeneities would account for
a correction to the off--proton scattering, which, in principle,
could be included into $np$ scattering transition probability.
However, in view of the anticipated smallness of this effect,
we will neglect it.

\section{Kinetic equation}
Transport of charge, heat or momentum is limited by scattering
of the elementary excitations (quasiparticles) off localized
protons and by their mutual
collisions. Among the latter we neglect those caused
by weak $en$ electromagnetic interaction, which enables us to
study transport properties of electrons and neutrons
separately.
The distribution function
of electrons satisfies
the
standard time--independent Boltzmann equation, valid for ideal gases,
(e.g.\ ref.\ \cite{Z60}),
whereas
the
neutron distribution is governed by
the
time--independent
Landau equation, which takes into account a self--consistent
mean field depending on the quasiparticle distribution itself.
However,
a proper linearization (e.g.\ ref.\ \cite{BP91}) reveals that
the quantity of real significance for
the
transport processes
(the deviation of the distribution function from the local
equilibrium one) satisfies the linearized Boltzmann equation,
which allows a unified treatment
for
electrons and neutrons.
We have
\begin{equation}
       {\bf v}_1 \,\, \partial_{\bf r} f_{{\bf k}_1} = I_{ii} + I_{ip},
\label{BEke}
\end{equation}
\begin{equation}
       e {\bf E} \,\, \partial_{{\bf k}_1} f_{{\bf k}_1} = I_{ee} + I_{ep},
\label{BEs}
\end{equation}
where $f_{{\bf k}_1}$ is the distribution function in question, $i=\{n,e\}$,
{\bf E} is an external electric field, ${\bf v}_1$ is the velocity of
quasiparticles, which is assumed to be independent of coordinates,
and ${\bf k}_1$ is the quasiparticle wavevector.
On the right hand sides of these equations we have integrals of
particle--particle and particle--proton collisions which can be
written as
\begin{eqnarray}
          I_{ii} &=& {(2 \pi)^4 \over (2 \pi)^9}
          \int {\rm d} {\bf k}_2 {\rm d} {\bf k}'_1 {\rm d} {\bf k}'_2
          \,\, \delta ({\cal E}'  - {\cal E}) \,
          \delta ({\bf K}'  - {\bf K}) \,\,
          {\cal L}_{ii} \,\,\,
          \frac{1}{2} \sum_{\sigma_2 \sigma'_1 \sigma'_2}
                      |A_{ii}|^2,
\label{Iii} \\
          I_{ip} &=& {2 \pi n_p \over (2 \pi)^3}
          \int {\rm d} {\bf k}'_1
          \,\, \delta (\varepsilon'_1 - \varepsilon_1) \,\,
          \frac{1}{2} \sum_{\sigma_p \sigma'_1 \sigma'_p}
                      |A_{ip}|^2
\nonumber \\
          &\times&
          \left[f_{{\bf k}'_1} (1-f_{{\bf k}_1}) -
                f_{{\bf k}_1} (1-f_{{\bf k}'_1}) \right] .
\label{Iip}
\end{eqnarray}
In this case,
non--primed and primed variables correspond to particles before and after
a collision, respectively,
$A_{ii}$ and $A_{ip}$ are
the transition amplitudes of the $ii$ and $ip$ scattering processes,
containing
all exchange contributions,
${\cal L}_{ii}$ is the standard ``two fermion'' Pauli factor,
and $n_p$ is the density of localized protons.
In the first integral the
factor 1/2 serves to avoid double
counting of the final states in a collision event. In the second
one the same factor accounts for the fact that the density
$n_p$ already includes protons with both possible
spin orientations.

It is useful to cast the spin sums in the preceding equations
in a more symmetric form:
\begin{equation}
          \frac{1}{2} \sum_{\sigma_2 \sigma'_1 \sigma'_2}
          |A_{ii}|^2 =
          \frac{1}{4} \sum_{\sigma_1 \sigma_2 \sigma'_1 \sigma'_2}
          |A_{ii}|^2 \equiv Q_{ii},
\label{Q-def}
\end{equation}
with analogous expression for $Q_{ip}$.

If the distribution functions are known,
the kinetic coefficients
can be found by calculating
the currents induced in the system by the spatial gradients or
the external field.
We have
\begin{eqnarray}
        {\bf j}_{\rm T} &=& {2 \over (2 \pi)^3} \int {\rm d} {\bf k} \,\,
                            f_{\bf k} \, (\varepsilon - \mu) \, {\bf v} =
                            - \kappa {\bf \nabla} T,
\label{jT} \\
        {\bf j}_{\rm e} &=& {2 \over (2 \pi)^3} \int {\rm d} {\bf k} \,\,
                            f_{\bf k} \, e \, {\bf v} = \sigma {\bf E},
\label{je} \\
      - \sigma_{\rm xy} &=& {2 \over (2 \pi)^3} \int {\rm d} {\bf k} \,\,
                            f_{\bf k} \, k_{\rm x} \, v_{\rm y} = - \eta
                            {\partial u_{\rm x} \over \partial y},
\label{sigma-XY}
\end{eqnarray}
where $\kappa$ is the thermal conductivity, $\sigma$ is the
electrical conductivity,
$\eta$ is the shear viscosity,
and {\bf u} is a bulk
velocity, which is assumed to be directed along the $x$--axis
and to be dependent only on the $y$ coordinate (thus satisfying
${\rm div}~{\bf u} = 0$).
Further, ${\bf j}_{\rm T}$, ${\bf j}_{\rm e}$, and $\sigma_{\rm xy}$
are, respectively, the heat flux, the charge flux and
the $xy$--component of the dissipative part of the stress tensor.

In the presence of the spatial gradients we approximate the true
distribution functions on the left hand side of the Boltzmann
equation (\ref{BEke}) by the local equilibrium (le)
distributions as
\begin{eqnarray}
        f^{\rm le}_{\kappa} (\varepsilon) &=&
                                \left[ 1 + \exp{
                               {\varepsilon-\mu \over T({\bf r})}}
                                \right]^{-1},
\label{lhs-k} \\
          f^{\rm le}_{\eta} (\varepsilon) &=&
                               \left[ 1 + \exp{
                               {\varepsilon-\mu-{\bf k}{\bf u}({\bf r})
                               \over T}} \right]^{-1}.
\label{lhs-e}
\end{eqnarray}
Simple estimates \cite{BP91} show that it is a plausible
approximation provided the typical scales of the gradients of temperature
or bulk velocity are large compared to the
quasiparticle
mean free path $l$.
On the left hand side of Eq.\ (\ref{BEs}) we use the global
equilibrium distribution $f^0(\varepsilon)$. Corresponding
condition of applicability
reads $|eEl| \ll T$.

On the right hand side we adopt the following decomposition of
the true distribution functions in terms of the $f^0$:
\begin{equation}
         f_{\kappa, \sigma, \eta} = f^0 \,+ \, {f^0 \,(1-f^0) \over T}
         \Phi_{\kappa, \sigma, \eta}.
\label{Phi}
\end{equation}
In this case $\Phi_{\alpha}$ is an unknown function of energy
and angular variables, which represents a nonequilibrium
correction to the distribution function.
It is clear that the nonequilibrium term should be tied to
the
Fermi surface.

The dependences on
angular and energy variables can be
separated and the functions $\Phi$ can be looked for in the form
\begin{eqnarray}
           \Phi_\kappa &=& - \tau_{ii} \Psi_\kappa(x) \,
                           {\bf v} {\bf \nabla} T, \,\,\,\,\,
\nonumber \\
           \Phi_\sigma &=& \tau_{ii} \Psi_\sigma(x) \,
                           {\bf v} {\bf E} e, \,\,\,\,\,
\nonumber \\
           \Phi_\eta &=& -\tau_{ii} \Psi_\eta(x) \,
                           {\bf v} {\bf \nabla} ({\bf k}{\bf u}).
\label{psi-def}
\end{eqnarray}
In this case a dimensionless function $\Psi_{\alpha}$ represents
a dependence on energy [$x=(\varepsilon - \mu)/T$],
while
angular dependences
are uniquely determined by the left hand sides of Eqs.\ (\ref{BEke}),
(\ref{BEs}); $\tau_{ii}$
is a constant, which has the meaning of
a typical time between the collisions of identical particles.

Omitting
standard details which include the linearization of the
collision integrals Eqs.\ (\ref{Iii}), (\ref{Iip}) with respect to
the small nonequilibrium corrections
(\ref{Phi})
and integration of the momentum
and energy conserving delta functions,
we arrive at the one--dimensional equation for the functions
$\Psi_\alpha$:
\begin{eqnarray}
    g_\alpha (x_1) f^0_1 (1-f^0_1) &=& {x^2_1+\pi^2 \over 2}
    f^0_1 (1-f^0_1) \Psi_\alpha(x_1)
\nonumber \\
    &+& \lambda_\alpha \int^{\infty}_{-\infty}
    {\rm d} x_2 \,\, f^0_1 f^0_2 {x_1+x_2 \over
    1 - {\rm e}^{-x_1-x_2}} \Psi_\alpha(x_2)
\nonumber \\
                       &+& {\tau_{ii} \over \tau_{ip}} \,
                       f^0_1 (1-f^0_1) \, (1-\omega_\alpha) \,
                       \Psi_\alpha(x_1),
\label{Psieq}
\end{eqnarray}
where $\alpha$ runs over $\kappa, \sigma$, and $\eta$,
$g_\kappa (x_1) = x_1$,
and
$g_{\sigma, \eta} = 1$.
The expression for the characteristic collisional time is
given by
\begin{equation}
       \tau_{ii} = {8 \pi^3 \over m^{* 3}_i T^2
                        \langle Q_{ii} \rangle},
\label{tauii}
\end{equation}
where the angle brackets denote an
angular
averaging of the form
\begin{equation}
       \langle Q_{ii} \rangle = \int_{(4 \pi)}
       {{\rm d} \theta {\rm d} \phi \over 4 \pi} \,\,
       {\sin{\theta} \over \cos{\theta/2}} \,\,
       Q_{ii} (\theta, \phi).
\label{Qii}
\end{equation}
In this case, the angles $\theta$ and $\phi$ specify a collision
of identical particles in the Abrikosov--Khalatnikov
frame of reference. Namely, $\theta$ is an angle between
momenta of first and second incident particles, and $\phi$
is an angle between planes, containing
two incident
and two final momenta. The other quantities characterizing
$ii$--type collisions are
\begin{eqnarray}
      \lambda_\kappa &=& {1 \over \langle Q_{ii} \rangle} \,
      \langle Q_{ii} (\theta, \phi) \, (1 + 2 \cos{\theta}) \rangle,
\nonumber \\
      \lambda_\eta &=& {1 \over \langle Q_{ii} \rangle} \,
      \langle Q_{ii} (\theta, \phi) \,
      (3 \sin^4{\frac{\theta}{2}} \,\, \sin^2{\phi} - 1) \rangle,
\label{lambda}
\end{eqnarray}
and $\lambda_\sigma = -1$. The latter relation corresponds to the
fact that $ee$ collisions do not limit the transport of electric
charge.

A characteristic time between collisions of a particle with localized
protons reads
\begin{equation}
      \tau_{ip} = {2 \over n_p m^*_i
      k_{Fi} \langle Q_{ip} \rangle},
\label{tauip}
\end{equation}
where in the case of $ip$--collisions the angle brackets stand for
an averaging according to
\begin{equation}
      \langle Q_{ip} \rangle = \frac{1}{\pi} \,\,
      \int^{\pi}_0 {\rm d} \chi  \, \sin{\chi} \, Q_{ip}(\chi).
\label{Qip}
\end{equation}
$\chi$ is a scattering angle of a particle colliding with a proton.
Finally, the efficiency of $ip$--collisions is described by the
quantities $\omega_\alpha$ defined as:
\begin{eqnarray}
       \omega_\kappa &=& \omega_\sigma = {1 \over \langle Q_{ip} \rangle}
       \,\, \langle Q_{ip} (\chi) \cos{\chi} \rangle, \,\,\,\,\,
\nonumber \\
       \omega_\eta &=& {1 \over \langle Q_{ip} \rangle}
       \,\, \langle Q_{ip} (\chi) \, \frac{1}{2} \,
       (3 \cos^2{\chi} - 1) \rangle.
\label{omega}
\end{eqnarray}

The Eqs.\ (\ref{Psieq}) can be solved exactly if either
particle--proton collisions dominate
($\tau_{ii} \gg \tau_{ip}$) or in the opposite case.
In the former situation we have simply
$\Psi_\alpha(x) = g_\alpha(x) \tau_{ip}
[\tau_{ii} (1-\omega_\alpha)]^{-1}$.
If, in contrast,
the $ip$--collisions
are negligible the exact solution is nontrivial and can be
obtained with the aid of a method
developed
by Brooker and Sykes
\cite{BS68} and by H{\o}jgaard Jensen {\it et al.} \cite{HJetal68}.
Finally, it is possible
to find variational solutions \cite{Z60}
for the functions $\Psi_\alpha$ at arbitrary value of the ratio
$\tau_{ii}/\tau_{ip}$. This is done by assuming
specific dependences of these functions on $x$
($\Psi_\kappa \propto x, \Psi_{\sigma,\eta} = {\rm const}$),
which are consistent with symmetries of Eq.\ (\ref{Psieq}),
and by subsequent determination of the unknown coefficients
on integration of the equation over $x_1$ within infinite
limits (the thermal conduction equation must be, in addition, multiplied
by $x_1$).
The resulting expressions are:
\begin{eqnarray}
      \Psi_\kappa (x) &=& \left [{2 \over 5} \, \pi^2 \,
                             (3 - \lambda_{\kappa})
                             \, + \, {\tau_{ii} \over {\tau}_{ip}}
                             \, (1-\omega_\kappa) \right]^{-1} \, x,
\label{psi-k} \\
      \Psi_{\sigma,\eta} (x) &=& \left[ {2 \over 3} \, \pi^2 \,
                             (1 + \lambda_{\sigma,\eta})
                             \, + \, {\tau_{ii} \over {\tau}_{ip}}
                             \, (1-\omega_{\sigma,\eta}) \right]^{-1}.
\label{psi-se}
\end{eqnarray}
When
the functions $\Psi_\alpha$ are known the calculation
of
the
transport coefficients
becomes
straightforward. We insert
Eqs.\ (\ref{Phi}), (\ref{psi-def}), (\ref{psi-k}), and
(\ref{psi-se}) into Eqs.\ (\ref{jT}) -- (\ref{sigma-XY})
and
find
\begin{eqnarray}
           \kappa_i &=&
           {\pi^2 \, n_i \, T \over 3 \, m^*_i \,
           (\nu_{\kappa ii} + \nu_{\kappa {ip}})},
\label{kappa-var} \\
            \sigma_e &=&
            {e^2 \, n_e \over m^*_e \nu_{\sigma {ep}}},
\label{sigma-var} \\
            \eta_i &=&
            {m^*_i \, n_i \, v^2_{Fi} \over 5 \,
            (\nu_{\eta ii} + \nu_{\eta {ip}})}.
\label{eta-var}
\end{eqnarray}
In this case $\nu_{\alpha ii}$ are the effective frequencies of
particle--particle collisions:
\begin{equation}
      \nu_{\kappa ii} = {m^{\ast 3}_i T^2 \over 20 \pi}
       \,\, \langle Q_{ii} \rangle \, (3-\lambda_\kappa)
      = \frac{3}{5} \, \nu_{\eta ii} \,
      {3-\lambda_\kappa \over 1+\lambda_\eta},
\label{nuii}
\end{equation}
while $\nu_{\alpha {ip}}$ are those of particle--proton collisions:
\begin{eqnarray}
    \nu_{\kappa {ip}} &=& \frac{1}{2} n_p m^*_i k_{Fi}
                             \, \langle Q_{ip} \rangle \,
                            (1-\omega_\kappa) =
    \nu_{\eta {ip}} \, {1-\omega_\kappa \over 1-\omega_\eta},
\nonumber \\
    \nu_{\sigma {ep}} &=& \nu_{\kappa {ep}}.
\label{nuip}
\end{eqnarray}

The above variational expressions for
the
transport coefficients
are exact in the limit when $ip$--collisions dominate.
However, in the opposite limit the variational method
appears to be not very accurate (most notably for
the
thermal conductivity).
The exact formulae of
refs.\ \cite{BS68,HJetal68} differs from ours by some factors,
depending on the parameters $\lambda$. We will come back to those
factors and correct our equations in Section 5.
Now we turn to
a calculation of
the
angular integrals which appear in the
above equations for the effective frequencies.

\section{Scattering probabilities}
%

\subsection{Electrons}
Let us summarize briefly the main assumptions that facilitate
the angular averaging for electrons in
Eqs.\ (\ref{Qii}) -- (\ref{lambda}), and (\ref{Qip}) -- (\ref{omega}).
The electron--electron scattering is described adequately by
a Coulomb potential,
screened by modifications of
the
charge density induced in the vicinity
of each charge. This screening, being purely electron
(as the protons are fixed and in no way respond to
a small
perturbation), is represented by a dielectric function
$\epsilon(\omega,{\bf q})$ [$(\omega,{\bf q})$ -- is a 4--momentum
transfer in a collision event] and results mainly in a strong
suppression of
the
collision probability when the momentum transfer
$q$ is smaller than
$q_{\rm TF} = 2 \sqrt{\alpha / \pi v_{Fe}} \, k_{Fe}$ --
the Thomas--Fermi wave number
(here $\alpha$ is the fine--structure constant).
When $\omega \ll q v$ (where $v$ is
a typical velocity of electrons) the screening is static and
the dielectric function may be approximated by
\cite{J62}
\begin{equation}
      \epsilon(0,{\bf q}) = 1 + {\alpha \over \pi u^2} \left(
      \frac{2}{3} - {1-3 u^2 \over 6 u} \, \ln{\left|\displaystyle{{1-u \over
      1+u}}\right|} - {u^2 \over 3} \,
      \ln{\left|\displaystyle{{1-u^2 \over u^2}}\right|}
      \right),
\label{epsfull}
\end{equation}
where $u=q / 2 k_{Fe}$, and we took into account that electrons are
ultrarelativistic. In practice, the required smallness of energy
transfers for a degenerate system puts limit on its temperature:
$T \ll q v_{Fe} \sim q_{\rm TF} v_{Fe} = \sqrt{3} T_{pe}$,
where $T_{pe}$ is an electron plasma temperature. In terms of
the degeneracy parameter for
the
ultrarelativistic system this means
that
$T \ll 0.1 \varepsilon_{Fe}$ --- the condition we assume to be met.

Furthermore, one can easily verify that it is reasonable to approximate
the right hand side of the Eq.\ (\ref{epsfull}) for any $q< 2 k_{Fe}$,
by the formula
\begin{equation}
        \epsilon(0,{\bf q}) = 1 + {q^2_{\rm TF} \over q^2},
\label{epsappr}
\end{equation}
valid strictly only for small momentum transfers $q \ll 2 k_{Fe}$.
Even with the latter simplification the integrations in
Eqs.\ (\ref{Qii}), and (\ref{lambda}) remain cumbersome.
However, for ultrarelativistic particles the ratio
$y = q_{\rm TF} / 2 k_{Fe}$ is a small number
and, for practical
applications, it is sufficient
to retain only the lowest order terms of expansions of the
sought--for quantities in $y$. In this way we obtain
\begin{equation}
   \langle Q_{ee} \rangle  = {3 \pi^3 \over 2} \, {e^4
   \over \varepsilon^4_{Fe} y^3}, \,\,\,\,\,\,\,
   \lambda_\kappa = - \frac{1}{3}, \,\,\,\,\,\,
   \lambda_\eta = -1 + 10 y^2.
\label{ee-res}
\end{equation}
The resulting expressions for partial electron--electron thermal
conductivity and viscosity coincide with those obtained by
other
authors (e.g.\ refs.\ \cite{FI76,GY95}) in the
ultrarelativistic limit.

The treatment of the electron--proton collisions is somewhat
simpler. First of all, here,
the electron screening is always
static (unless, of course, the temperature is large
enough to excite the protons at their sites). The situation
with proton--proton correlations is not quite certain.
According to the arguments given in Section 2, we assume
that the proton system is completely disordered.
Under such conditions the problem of $ep$ scattering
is very similar to the scattering of electrons off impurities
which was studied e.g.\ by Flowers and Itoh \cite{FI76}. In this case the
integrations can be performed exactly at any degree of
electron relativism and at any $y$, but again only
the lowest order terms in $y$ in Eqs.\ (\ref{Qip}), and (\ref{omega})
are needed. The exact expressions can be found in
\cite{FI76}, however, the formula for viscosity
obtained
in this
work is slightly inaccurate. For this reason we give here both
exact and approximate expressions for
the shear
viscosity and an approximate
formula for
the
thermal and electrical conductivities:
\begin{equation}
     \langle Q_{ep} \rangle = {2 \pi e^4 \over
     k^4_{Fe} y^2}, \,\,\,\,\,\,\,
     \omega_\kappa = 1 +  4 y^2 (\ln{y}+1), \,\,\,\,\,\,
     \omega_\eta = 1 + 12 y^2 (\ln{y}+1.25).
\label{ep-appr}
\end{equation}
The exact expression for
the
viscosity reads
\begin{eqnarray}
    \langle Q_{ep} \rangle \, (1-\omega_\eta) &=&
    {12 \pi e^4 \over k^4_{Fe}} \left\{
    [1 + 2 y^2 (1 + \beta^2) + 3 y^4 \beta^2] \,
    \ln{\displaystyle{{1+y^2 \over y^2}}} \right.
\nonumber \\
    &-& \left. 2 - {\beta^2 \over 2}
    - 3 y^2 \beta^2 \right\},
\end{eqnarray}
where $\beta = v_{Fe}/c$.

Finally, we note that the results obtained for $ep$ scattering,
remain
unchanged,
if the protons are fully spin polarized.

\subsection{Neutrons}
Contributions of neutrons and protons to the thermal conductivity
and viscosity for the standard non--localized model of matter
in neutron star cores were studied by Flowers and Itoh \cite{FI79}.
In the case of nucleon contributions a great deal
of uncertainty is related
to the description of the scattering processes.
It is impossible to make use of
the Landau theory for $nn$ scattering, since at present
we do not know values of the momentum dependent quasiparticle
amplitudes in the density range of interest. Yet even worse is our
understanding of scattering of a neutron quasiparticle
at the Fermi surface off a localized proton.

Another approach, fundamentally cruder, but able to supply
us with the desired physical input, consists of a neglect of
an influence of many--body effects on the scattering amplitudes.
This means
that we base our consideration on the data on nucleon--nucleon (NN)
scattering {\it in vacuum}. An obvious disadvantage of this approach
is an inappropriate treatment of the
short--range
repulsive part of the
NN interaction: being to some extent screened in
a collision of quasiparticles, it is well sampled in a collision
of bare nucleons. Therefore, such a method is expected to overestimate
the role of the collisional processes in limiting the neutron
transport, thus underestimating the transport coefficients.
Although, following this second way, we hope that it must
yield correct order--of--magnitude estimate of
the neutron transport coefficients, we warn the reader that
our results can be a few times less than the actual values
of the quantities in question.

Within this framework the transition probabilities for
$nn$ collisions at any degree of relativism
are easily reconstructed
from vacuum differential cross sections in the center--of--mass
(cm) reference frame, namely:
\begin{equation}
         Q_{nn} (\theta, \phi) = {16 \pi^2 \hbar^4 \over
         m^2_N + k^2_{Fn} \sin^2{\theta/2}}
         {{\rm d} \sigma_{nn} \over {\rm d} \Omega_{\rm cm}}
         (E_{\rm lab},\phi_{\rm cm}),
\label{cross-nn}
\end{equation}
where
\begin{equation}
    E_{\rm lab} = {k^2_{Fn} \over m_N} (1- {\rm cos} \theta),
    \,\,\,\,\, {\rm and} \,\,\,\,\,
    \phi_{\rm cm} = \phi,
\label{nnaux}
\end{equation}
are the collision energy in the laboratory (lab) reference frame and
the cm scattering angle.

The situation is more problematical for the case of $np$ collisions.
Under the condition that neutrons are not relativistic,
the transition probability for the scattering of a neutron off
a localized proton (which can be thought of as a neutron scattering
off an external field describing by the same $np$ potential)
can be again derived from the differential cross sections of
$np$ scattering in vacuum as
\begin{equation}
         Q_{np} (\theta) = {16 \pi^2 \hbar^4 \over
         m^2_N} {{\rm d} \sigma_{np} \over {\rm d} \Omega_{\rm cm}}
         (E_{\rm lab}, \phi_{\rm cm}),
\label{cross-np0}
\end{equation}
where
\begin{equation}
          E_{\rm lab} = {2 k^2_{Fn} \over m_N},
          \,\,\,\,\, {\rm and} \,\,\,\,
          \phi_{\rm cm} = \theta.
\label{npaux}
\end{equation}
However, with growing relativism, such a procedure ceases to be adequate,
as the scattering off an instantaneous external potential
becomes different from the scattering of two particles
(the case studied in a laboratory).
But for the densities considered ($\le 5 n_0$)
the neutrons are only moderately relativistic, and,
consequently, we may adopt the above formalism with
the natural modification of the Eq.\ (\ref{cross-np0}):
\begin{equation}
         Q_{np} (\theta) = {16 \pi^2 \hbar^4 \over
         m^2_N + k_{Fn}^2}
         {{\rm d} \sigma_{np} \over {\rm d} \Omega_{\rm cm}}
         (E_{\rm lab}, \phi_{\rm cm}).
\label{cross-np}
\end{equation}

We have calculated the required angular averages of the transition
probabilities [those appearing in the Eqs.\ (\ref{nuii}), and
(\ref{nuip})] using the tables of vacuum cross sections
in the cm reference frame as functions of
lab energy (in the range from 10 to 700 MeV) and cm scattering angle.
The tables themselves for $pp$ and $np$ scattering were
obtained by using the partial wave solution WI96 available
in the SAID database \cite{Aetal83} developed at the
Virginia Polytechnic Institute by R.A.\ Arndt with collaborators.
To construct the $nn$ cross sections from the $pp$ ones
the following procedure was used. First of all we subtracted
from the table values the well known values of the Coulomb
cross sections. This gave us reasonable estimates of the
$nn$ cross sections at larger angles but made the tables
inapplicable at smaller angles (e.g.\ at energies less than
400 MeV the table values at 5$^{\rm o}$ became negative,
indicating the importance of the interference terms).
It was tempting then to extrapolate smoothly the values
of the $nn$ cross sections at larger angles ($\ga 20^{\rm o} - 40^{\rm o}$
depending on energy) to the domain of small angles,
which yielded a reasonable estimate of the $nn$ cross sections
over the entire angle range. Finally, at zero energy we
have used the value of 3030 mb that
followed from the theory of $nn$ scattering length
(e.g.\ ref.\ \cite{PB75}).

The calculations were done
for the values of neutron Fermi wave vector $k_{Fn}$
from 1.1 to 2.9 fm$^{-1}$ ($n_n$ from 0.5 to 5 $n_0$).
To interpolate between the neighbour
nodes of the tables
of the cross sections
the bilinear interpolation was used.
The results of our calculations are fitted with the mean and
maximum errors of the fits less than 1.5 \% by the following
analytical expressions:
\begin{eqnarray}
    S_{\kappa n} &=& {m^2_N \,\, \langle Q_{nn} \rangle \,
    (3-\lambda_\kappa) \over 256 \, \pi \hbar^4 \, 1 {\rm mb}} =
    {0.3833 \over z_n^3 \sqrt{z_n}} + 3.652 z_n^{0.4},
\nonumber \\
    S_{\eta n} &=& {m^2_N \,\, \langle Q_{nn} \rangle \,
    (1+\lambda_\eta) \over 192 \, \pi \hbar^4 \, 1 {\rm mb}} =
    {0.1152 \over z_n^3} + {3.965 z_n \over 2.499 +
    z_n^6 \sqrt{z_n}},
\nonumber \\
     S_{\kappa p} &=& {m^2_N \,\, \langle Q_{np} \rangle \,
     (1-\omega_\kappa) \over 16 \, \pi \hbar^4 \, 1 {\rm mb}} =
     {1.833 \over z^2_n} + {1.430 z^2_n \over 0.3958 + z^8_n},
\nonumber \\
     S_{\eta p} &=& {m^2_N \,\, \langle Q_{np} \rangle \,
     (1-\omega_\eta) \over 16 \, \pi \hbar^4 \, 1 {\rm mb}} =
     {0.5663 \over z^3_n} + {4.545 z_n \over 1.276 + z^6_n}.
\label{nfits}
\end{eqnarray}
In all these expressions the quantity $z_n$ is the
neutron Fermi wave vector in units of 2.666 fm$^{-1}$
(corresponding to density 4$n_0$).

\section{Practical formulae}
Let us summarize the results, derived in the previous sections,
and present them in the form useful for practical applications.
First of all, we note that
the
variational solutions obtained could
be corrected to yield the exact asymptotes in the limit
when particle--particle collisions dominate. This is done customarily
by multiplying the
partial particle--particle
variational transport coefficients
by well--known factors $C_\alpha$ depending on the parameters
$\lambda_\alpha$. In the case of electrons $C_\kappa(-1/3) = 1.3$,
while $C_\eta(-1+y^2) \approx 1$, i.e.\ the variational solution
for
the shear
viscosity is exact for ultrarelativistic electrons.
For neutrons the coefficients $C_\alpha$ are weakly varying
functions of density. For the considered density range it is
a very good approximation to adopt fixed values of the correction
factors $C_\kappa = 1.2$ and $C_\eta = 1.05$.

Bringing together Eqs.\ (\ref{kappa-var}) -- (\ref{nuip}),
(\ref{ee-res}), (\ref{ep-appr}), and (\ref{nfits}), and
incorporating the above factors, we may write for
the
thermal conductivity:
\begin{eqnarray}
    \kappa^{-1}_i &=& \kappa^{-1}_{ii} + \kappa^{-1}_{ip},
\nonumber \\
    \kappa_{ee} &=& 2.2 \cdot 10^{23} \,\,\,
    \left( {n \over 4 n_0} \cdot {x_e \over 0.01} \right) \, {1 \over T_8}
    \,\,\,\,\, {\rm ergs \,\,\,\, s^{-1} cm^{-1} K^{-1}},
\nonumber \\
    \kappa_{ep} &=& 2.5 \cdot 10^{19} \,\,\,
    \left( {n \over 4 n_0} \cdot {x_e \over 0.01} \right)^{1/3} \, T_8
    \,\,\,\,\, {\rm ergs \,\,\,\, s^{-1} cm^{-1} K^{-1}},
\nonumber \\
    \kappa_{nn} &=& 8.3 \cdot 10^{23} \,\,\,
    \left( {m_N \over m^*_n} \right)^4 \,
    \left( {n_n \over 4 n_0} \right) \, {1 \over T_8} \,
    {1 \over S_{\kappa n}}
    \,\,\,\,\, {\rm ergs \,\,\,\, s^{-1} cm^{-1} K^{-1}},
\nonumber \\
    \kappa_{np} &=& 8.9 \cdot 10^{17} \,\,\,
    \left( {m_N \over m^*_n}\right)^2 \,
    \left( {n_n \over 4 n_0}\right)^{2/3} \,
\nonumber \\
    &\times&
    \left( {n \over 4 n_0} \cdot {x_p \over 0.01} \right)^{-1} \,
    {T_8 \over S_{\kappa p}}
    \,\,\,\,\, {\rm ergs \,\,\,\, s^{-1} cm^{-1} K^{-1}},
\label{thcond-res}
\end{eqnarray}
for
the
electrical conductivity:
\begin{equation}
      \sigma_e = \sigma_{ep} = 9.2 \cdot 10^{23} \,\,\,
      \left( {n \over 4 n_0} \cdot {x_e \over 0.01} \right)^{1/3}
      \,\, {\rm s}^{-1},
\label{econd-res}
\end{equation}
and for
the shear
viscosity:
\begin{eqnarray}
    \eta^{-1}_i &=& \eta^{-1}_{ii} + \eta^{-1}_{ip},
\nonumber \\
    \eta_{ee} &=& 1.7 \cdot 10^{19} \,\,\,
    \left( {n \over 4 n_0} \cdot {x_e \over 0.01} \right)^{5/3} \,
    {1 \over T^2_8}
    \,\,\,\,\, {\rm g \,\, cm^{-1} s^{-1}},
\nonumber \\
    \eta_{ep} &=& 1.1 \cdot 10^{13} \,\,\,
    \left( {n \over 4 n_0} \cdot {x_e \over 0.01} \right)
    \,\,\,\,\, {\rm g \,\, cm^{-1} s^{-1}},
\nonumber \\
   \eta_{nn} &=& 1.5 \cdot 10^{19} \,\,\,
   \left( {m_N \over m^*_n} \right)^4 \,
   \left( {n_n \over 4 n_0} \right)^{5/3} \,
   {1 \over T^2_8} \, {1 \over S_{\eta n}}
    \,\,\,\,\, {\rm g \,\, cm^{-1} s^{-1}},
\nonumber \\
   \eta_{np} &=& 2.2 \cdot 10^{13} \,\,\,
    \left( {m_N \over m^*_n} \right)^2 \,
    \left( {n_n \over 4 n_0}\right)^{4/3} \,
\nonumber \\
     &\times&
    \left( {n \over 4 n_0} \cdot {x_p \over 0.01} \right)^{-1} \,
    {1 \over S_{\eta p}}
    \,\,\,\,\, {\rm g \,\, cm^{-1} s^{-1}}.
\label{visc-res}
\end{eqnarray}

Let us remind the conditions of applicability of the above
expressions. Both electrons and neutrons must be strongly degenerate.
The neglect of the dynamical screening effect is based on
a more stringent condition for electrons: $T \ll 0.1 \varepsilon_{Fe}$.
The electron number density must ensure their
ultrarelativism. Finally, the density of neutrons is restricted
to the range from 0.5 to 5 $n_0$.

\section{Neutrino losses}
In this section we study two processes, contributing to neutrino
cooling, which are specific for neutron star matter with localized
protons. The first of these processes is the
neutrino--antineutrino pair bremsstrahlung due to electron scattering
off localized protons. The second process is the strong--interaction analog
of the first one: it is the neutrino--antineutrino pair bremsstrahlung
accompanying scattering of neutrons off protons localized in neutron
medium.
The temperature
dependence of both these processes is
$T^6$,
that is the same as for the direct URCA process. Therefore,
the localization of protons may imply non--standard
(accelerated) cooling of a neutron star. Below, we will
estimate the rate of the energy losses in these processes
under some model assumptions.

{\it (a) Electrons.}
In the case of electrons we can use the results of ref.\ \cite{HKY96}.
This paper was
concerned with the neutrino--pair
bremsstrahlung due to electron--nucleus scattering in the
liquid phase of the neutron star crusts. The authors derived
the  following general expression for the energy loss rate
(Eq.\ (8) of \cite{HKY96})
\begin{equation}
       Q_{\rm Brem}^e = {8 \pi G_{\rm F}^2 Z^2 e^4 C^2_+
                        \over 567 \hbar^9 c^8} (k_B T)^6 n_i L
                        \,\,\,\,\, {\rm ergs~s}^{-1}~{\rm cm}^{-3},
\label{Qegen}
\end{equation}
where $G_F = 1.436 \cdot 10^{-49}$ ergs cm$^3$ is the
Fermi weak coupling constant, factor $C^2_+ = 1.675$ takes into account
the emission of $\nu_e, \nu_\mu$, and $\nu_\tau$, $Z$ is the nucleus charge,
$n_i$ is the number density of nuclei, and the quantity $L \approx 1$,
interpreted as a Coulomb logarithm, is a weakly varying
function of $Z$, $T$, and $n_i$. The authors derived
also the general formula for $L$ [their Eq.\ (19)], which takes into
account the nucleus electromagnetic formfactor,
the static structure factor of nuclei, static electron screening,
non--Born corrections and incorporates accurately the thermal
effects. They also proposed an analytical fitting formula
for $L$ [Eq.\ (25)]. In our case we have obviously $Z=1$
and $n_i = n_p$. The fit of \cite{HKY96} does not
apply here (the authors were interested in the crust
and considered $Z \ge 10$). Besides, in our
situation the temperature is much lower than the
screening momentum, {\it the low--temperature case} of \cite{HKY96}.
In this regime the thermal effects are not important and
$L$ is given by much simpler formula (21) of \cite{HKY96}.
In this expression we can neglect non--Born corrections and
nucleus formfactor as well as nuclei structure factor
(because our proton system is assumed to be fully disordered), and
approximate static dielectric electron screening function
by Eq.\ (\ref{epsappr}). Performing one--dimensional integration
for ultrarelativistic electrons, we obtain $L = 1.755$.
Inserting this value into Eq.\ (\ref{Qegen}) we obtain
\begin{equation}
   Q^{e-{\rm loc.p.}}_{\rm Brem} = 6.0 \cdot 10^{18} \, T_9^6 \,
                       \left( {n \over 4 n_0} \cdot {x_e \over 0.01} \right)
                        \,\,\,\,\, {\rm ergs~s}^{-1}~{\rm cm}^{-3},
\label{Qenum}
\end{equation}
where $T_9 \equiv T/10^9$~K.

{\it (b) Neutrons.}
In this subsection we will obtain an expression,
which will
enable us to estimate the energy loss rate due to
neutrino pair emission from neutron scattering off
a localized proton. To simplify the derivation we will,
first, regard the neutrons as fully non--relativistic,
and, second, will treat $np$--interaction in a very crude
manner, assuming it to be given by a contact spin--independent
potential, which will be treated
in the Born approximation.
The interaction strength $U$ will further be
made density--dependent
by fitting it at a given density (or, equivalently, at a given
collision energy) to the total vacuum $np$ cross--section,
known experimentally or theoretically (see below).
No correction of the $np$ scattering rate to account for the presence
of the medium (except for the exclusion principle in the initial
and final neutron states) will be made.

The process in question, in the formalism of
Feynman diagrams,
is described (to lowest  order) by two diagrams, (A) and (B).
In both cases, we consider
an initial neutron with 4--momentum $k = (\varepsilon, {\bf k})$.
In the case of diagram (A), this neutron first emits
a neutrino--antineutrino pair
with total 4--momentum $p = (\omega,{\bf p}) = k_1 + k_2$,
where $k_1 = (\omega_1,{\bf k}_1)$ and $k_2 = (\omega_2,{\bf k}_2)$
are, respectively, 4--momenta of neutrino
and antineutrino, then propagates with 4--momentum $k-p$
and finally interacts strongly with a localized proton,
absorbing 4--momentum
$q = (0,{\bf q})$ and ending in the final state with 4--momentum
$k' = (\varepsilon',{\bf k'})$. Diagram (B) corresponds to
the situation, in which neutron first interacts
strongly with a localized proton, then propagates with 4--momentum $k'+p$,
and, finally, emits a neutrino--antineutrino
pair of total 4--momentum $p$,  ending in the final
state $k'$.

We will use
nonrelativistic formalism (for neutrons) to
describe the weak-interaction vertex,
take non-relativistic neutron Green functions,
when dealing with the intermediate states,
and nonrelativistic spinors for neutron initial and final
states. Then the first order matrix element assumes the form
\begin{equation}
    M = M_A+M_B = - {i G_F U \over 2 \sqrt{2}} \,
        \chi'^\dagger (\delta_{\mu 0} - g_A \delta_{\mu i} \sigma_i) \chi \,
        [G(k-p) + G(k'+p)] \, l^\mu,
\label{MAB}
\end{equation}
where $g_A \approx 1.26$ is the axial
renormalization constant, greek indices
take the values 0, 1, 2, 3,
and latin indices take the values 1, 2, 3,
$\sigma_i$
are  the standard Pauli matrices, $\chi$ and $\chi'$
are the Pauli spinors, describing the initial and final
neutron states.
Further, $l^{\mu}$ is the neutrino neutral current given by
\begin{equation}
          l^\mu = \bar{u}_1 \gamma^{\mu} (1+\gamma^5) u_{-2},
\label{lmu}
\end{equation}
where $u_1 = u(k_1)$ and $u_{-2} = u(-k_2)$ are respectively
neutrino and antineutrino
bispinors, and bar means Dirac
conjugate.
In our problem the denominators of the neutron propagators $G$
cannot be zero. Hence, we can replace them by
the vacuum propagators, in spite of the fact,
that our process goes in the presence of neutron Fermi sea.
Nonrelativistic  neutron propagator takes
therefore the form
\begin{equation}
         G(\omega,{\bf k}) = {1 \over \omega - \varepsilon_{\bf k} + i0}.
\label{Green}
\end{equation}

The neutrino pair energy loss rate (in erg cm$^{-3}$ s$^{-1}$) is
given by
\begin{eqnarray}
    Q^{n-{\rm loc.p.}}_{\rm Brem}  &=& 3 n_p \,
            \int {{\rm d} {\bf k} \,{\rm d} {\bf k'} \,{\rm d} {\bf k}_1 \,
            {\rm d} {\bf k}_2 \, {\rm d} {\bf q} \over
            4 \omega_1 \omega_2 \, (2 \pi)^{15}} \, (2 \pi)^4 \,
            \delta^{(4)}(k+q-k'-k_1-k_2)
\nonumber \\
      &\times&  f(1-f') \, \omega \, \sum_{ss'} |M|^2,
\label{Qgen}
\end{eqnarray}
where $n_p$ is the number density of localized protons, $f = f(\varepsilon)$
and $f' = f(\varepsilon')$ are the neutron Fermi--Dirac distribution
functions, and the summation runs over final and initial
neutron spin states. A factor of 3 accounts for the emission of
$\nu_e, \nu_\mu$, and $\nu_\tau$ pairs.

Calculating the spin--summed  squared  matrix elements in
the straightforward manner and using the identity \cite{BLP82}
\begin{equation}
      \int {{\rm d}{\bf k}_1 \, {\rm d}{\bf k}_2 \over \omega_1 \omega_2}
       \, k_1^\mu k_2^\nu \delta^{(4)}(p-k_1-k_2) =
       {\pi \over 6} \, (p^2 g^{\mu \nu} + 2 p^\mu p^\nu),
\label{ident}
\end{equation}
we obtain the following expression for the emissivity
\begin{eqnarray}
   Q^{n-{\rm loc.p.}}_{\rm Brem}
 &=& {\pi G_F^2 U^2 n_p \over (2 \pi)^{11}}
    \int {\rm d}{\bf k} \, {\rm d}{\bf k'} \, {\rm d}{\bf p} \,
    f(1-f') \, \omega \, [3 g^2_A \omega^2 + (1-2 g^2_A) {\bf p}^2]
\nonumber \\
    &\times&  4 m^2_n
    \left( {1 \over 2 m_n \omega - 2 {\bf p k'} - {\bf p}^2 }
         - {1 \over 2 m_n \omega - 2 {\bf p k} + {\bf p}^2} \right)^2,
\label{Qful}
\end{eqnarray}
where we have replaced the integration over d{\bf q}
by that over  d{\bf p}.
 In this expression
we must integrate over the domain, where $\varepsilon' < \varepsilon$
and for $|{\bf p}| < \omega = \varepsilon - \varepsilon'$.
In the denominators we can safely omit the terms ${\bf p}^2/2 m_n$
which are negligible compared to $\omega$, and using
the fact that neutrons are nonrelativistic, we can expand
the denominators
to first order
in ${\bf p k'}/ m_n$ and ${\bf p k}/ m_n$.
Then we direct the $z$--axis of the spherical reference frame
along the vector {\bf k} and place the vector ${\bf k'}$ into
the $xz$--plane. Integrations over d{\bf p} and over angle
between vectors {\bf k} and ${\bf k'}$ yield the emissivity
in the form of two--dimensional ``Fermi'' integral
\begin{eqnarray}
   Q^{n-{\rm loc.p.}}_{\rm Brem}
        &=& {8 \pi G^2_F U^2 n_p \over 21 (2 \pi)^8} \,
        (1 + 2.2 g^2_A) \,
            \left( m^\ast_n \over m_n \right)^2 \,
\nonumber \\
        &\times&
          \int {\rm d} \varepsilon \, {\rm d} \varepsilon'
          |{\bf k}| |{\bf k'}| ({\bf k}^2 + {\bf k'}^2) f (1-f') \omega^4
\nonumber \\
          &=& {16 \pi G^2_F U^2 n_p \over 21 (2 \pi)^8} \,
            (1 + 2.2 g^2_A) \,
            \left( m^\ast_n \over m_n \right)^2 \,
              k_{Fn}^4 T^6
\nonumber \\
         &\times&  \int^{\infty}_{-\infty} {\rm d}x
              \int^{\infty}_0 {\rm d} \omega \,
              {\omega^4 \over (e^x+1)(e^{\omega-x}+1)},
\label{Q2d}
\end{eqnarray}
where we have taken into account that neutrons are strongly
degenerate and have pulled all the smooth functions
of the neutron momenta out of the integral at the Fermi surface.
The latter integral is standard and is equal to $(2 \pi)^6 / 504$.
Substituting this value into Eq.\ (\ref{Q2d}) we obtain
\begin{equation}
   Q^{n-{\rm loc.p.}}_{\rm Brem}
              = {G^2_F U^2 \over 2646 \pi} \,
              (1 + 2.2 g^2_A) \,
            \left( m^\ast_n \over m_n \right)^2 \,
                  n_p k_{Fn}^4 T^6.
\label{Qfin}
\end{equation}
The last step is to specify the value of $U^2$.
For the contact
spin--independent $np$ interaction $U^2$ gives differential
probability of elastic neutron scattering
off a localized proton (the same for all angles $\theta$).
Our approximation consists in  expressing  $U^2$ in terms
of the  total elastic cross section for $np$ scattering
in vacuum, using
Eqs.\ (\ref{cross-np0}, \ref{npaux}). Integrating over the
solid angle we have
\begin{equation}
      4 \pi U^2 = {16 \pi^2 \hbar^4 \over m^2_N} \,
                  \sigma_{np}^{\rm tot.el.}
                  \left( {2 k^2_{Fn} \over m_N} \right).
\label{Usgte}
\end{equation}
In this case, $\sigma_{np}^{\rm tot.el}$ is the total
elastic $np$ cross section which can be obtained from
the same theoretical model as in Section 4.2., WI96,
in the SAID database. We propose the following
fitting formula for $U^2$ in the energy range from 100 to 700 MeV:
\begin{equation}
      U^2 = {4 \pi \hbar^4 \,\,\, 1 \,\, {\rm mb} \over m^2_N} \,
            S^{\rm t}_{np}, \,\,\,\,\,
            {\rm where} \,\,\,\,\, S^{\rm t}_{np} =
            26.09 + {3.444 \over z^3_n},
\label{U2fit}
\end{equation}
and $z_n$ is defined below Eq.\ (\ref{nfits}). The mean error
of this expression is $\approx 1.2 \%$ and the maximum error
occurring at 700 MeV is $\approx 2.9 \%$.
Combining Eqs.\ (\ref{Qfin})  and (\ref{U2fit}), we obtain
the following numerical result:
\begin{eqnarray}
   Q^{n-{\rm loc.p.}}_{\rm Brem}
   &=& 1.3 \cdot 10^{21} \, S^{\rm t}_{np} \,
            \left( {n \over 4 n_0} \cdot {x_p \over 0.01} \right) \,
            \left( {n_n \over 4 n_0} \right)^{4/3} \,
            \left( m^\ast_n \over m_n \right)^2 \,
\nonumber \\
            &\times& T^6_9 \,\,\,\,\, {\rm ergs \, s^{-1} \, cm^{-3}}.
\label{Qnnum}
\end{eqnarray}
Medium effects can be expected to modify both
the magnitude and the density dependence of $U^2$, as compared
to that given by our simple prescription, Eq.\ (\ref{U2fit}).
Therefore, expression (\ref{Qnnum}) should be treated as a
rough estimate of $Q^{n-{\rm loc.p.}}_{\rm Brem}$.

\section{Astrophysical implications}
As one can see from the results obtained in the preceding sections,
presence of randomly distributed localized protons, removing
the $T^{-2}$ factor from
the
transport coefficients, leads at lower
temperatures to a dramatic decrease of $\kappa$, $\sigma$, and
$\eta$, as compared to the standard case, where protons form
a Fermi liquid. Let us discuss in some detail the astrophysical
implications of this effect for
neutron stars.

{\it (a) Thermal conductivity.}
In our case
transport of heat is dominated by electrons, which
scatter predominantly off localized protons, so that
$\kappa^{\rm loc.p.} \simeq \kappa_e^{\rm loc.p.} \simeq
\kappa_{ep}^{\rm loc.p.}$.
Let us compare this result with the values of $\kappa$
corresponding to the standard case. The latter are expressed by
the formulae derived in \cite{GY95}:
\begin{equation}
     \kappa_e =  {7.6 \cdot 10^{24} \over T_8}
                 \left({m_p\over m^*_p}\right)^{1/2}
                 \left({n_e \over n_0} \right)^{7/6}~
                  {\rm ergs~ cm^{-1} s^{-1} K^{-1}}
\label{GYns}
\end{equation}
for non--superfluid $p$, and
\begin{equation}
     \kappa_e =  {5.6 \cdot 10^{24} \over T_8}
                 \left({n_e \over n_0} \right)~
                 {\rm ergs~ cm^{-1} s^{-1} K^{-1}}
\label{GYs}
\end{equation}
for highly superfluid $p$,
the decisive difference between the two regimes being the presence
or absence of proton screening. At $n=4n_0$ and $x_e=0.01$ we thus get
\begin{equation}
          {\kappa_e \over \kappa^{\rm loc.p.}} \simeq
          {(5 \div 10) \cdot 10^3 \over T^2_8}~.
\label{kappas}
\end{equation}
The diffusive thermal conductivity due to neutrons is generally
on the order of or somewhat smaller than that of electrons.
(The actual ratio between them depends sensitively on
i) the fraction of electrons,
ii) the nucleon effective masses,
and iii) the strength of quasiparticle
interactions; in the non--superfluid regime $\kappa^{\rm loc.p.}_{nn}$
may serve as a good approximation of the neutron conductivity.)
Thus we can accept the value of $10^4 / T^2_8$ as a reasonable
estimate of the ratio $\kappa / \kappa^{\rm loc.p.}$.

Neutron stars are born as very hot objects which cool subsequently
due to neutrino losses from their interior. The cooling is accompanied
by thermal equilibration of the
stellar interior with a typical time scale,
determined by the size of the core $R_{\rm core}$,
the specific heat per unit volume $C$, and
the thermal conductivity of matter $\kappa$ as
\begin{equation}
       \tau_{\rm t.e.} \sim {R^2_{\rm core} C \over \kappa}.
\label{taute}
\end{equation}
For estimates, in the standard non--superfluid case
we can set $C = C_n$,
the specific heat of degenerate neutrons,
\begin{equation}
        C_n = 2.6 \cdot 10^{20} \left({m^\ast_n \over m_N} \right)
                   \left({n_n \over 4 n_0} \right)^{1/3} T_9
    \,\,\,\,\, {\rm ergs \,\,\,\, cm^{-3} K^{-1}},
\label{C_n}
\end{equation}
and approximate $\kappa$ by $\kappa_e$ (\ref{GYns}). Approximating
further all the effective masses by bare masses and assuming
a core of constant density ($n \simeq n_n = 4 n_0$)
and composition ($x_e = 0.01$) we obtain
$\tau_{\rm t.e.} \sim 460~T^2_9~(R_{\rm core} / 10~{\rm km})^2$ years.
If the core is superfluid we may use for $C$ the specific heat of
electrons $C_e$:
\begin{equation}
        C_e = 6.6 \cdot 10^{18} \left({n \over 4 n_0} \cdot
              {x_e \over 0.01} \right)^{2/3} T_9
    \,\,\,\,\, {\rm ergs \,\,\,\, cm^{-3} K^{-1}},
\label{C_e}
\end{equation}
and take  $\kappa$ from Eq.\ (\ref{GYs}). This gives us
$\tau_{\rm t.e.} \sim 10~T^2_9~(R_{\rm core} / 10~{\rm km})^2$ years
(the latter relation should be regarded as a lower bound,
as the neutron specific heat could decrease rather slowly
depending on the type of superfluidity \cite{LY94}).
The temperature dependence of both those expressions
is removed by the localization of protons. Multiplying
$\tau_{\rm t.e.}$ by the factor
$\kappa / \kappa^{\rm loc.p.}$ (\ref{kappas})
we obtain $\tau_{\rm t.e.}^{\rm loc.p.}$ equal to
$4.6 \cdot 10^4~ (R_{\rm core}^{\rm loc.p.} / 10~{\rm km})^2$ years and
$10^3~ (R_{\rm core}^{\rm loc.p.} / 10~{\rm km})^2$ years for
non--superfluid and strongly superfluid
cases, respectively.

The thermal conduction becomes important when the above
time scales are comparable or smaller than a characteristic
time of thermal evolution of matter due to
neutrino losses. Otherwise, the temperature of a given
element of matter is determined locally.
The neutrino cooling time scale could be estimated as
$CT/ \varepsilon_\nu$, where $\varepsilon_\nu$ is the total
neutrino emissivity. The temperature dependences
of $\varepsilon_\nu$ for various neutrino--emission processes
are usually very strong. In view of this,
the equilibration temperature for the case of proton localization
will not be much smaller, than that for the standard case;
in both cases the temperatures fall to the range
of 10$^8$ -- 10$^9$ K. However,
the time required to reach the
thermal equilibrium in a core, containing localized protons,
will be two orders of magnitude longer than that for a standard
liquid core. The latter conclusion
will not be changed by the appearance of
neutron superfluidity.

Another difference between the two cases
comes from the simple idea that the thermal equilibration time
in the localized protons case is temperature independent.
While in the standard case the core below $\sim 10^8$ K
could be treated as perfectly isothermal, it will still
take $\sim 10^3 - 10^4$ years to wash out any accidental
temperature inhomogeneity in the core with localized protons.

{\it (b) Electrical conductivity.}
The electrical conductivity
is relevant for the ohmic dissipation of
internal magnetic fields in neutron stars. In the standard case
of $npe$ matter with non--superfluid protons the charge transport
is dominated by ultrarelativistic electrons, scattering off protons,
which results in an electrical conductivity
$\sigma_{ep} = 2.1 \times 10^{31}~ (n_e/n_0)^{3/2} / T^2_8$ s$^{-1}$
\cite{BPP69}. Localization of protons prevents
the appearance of
a proton superconductor (as the localization temperature is rather high,
of the order of $10^{10} - 10^{11}$ K, which is commonly believed
to be higher than $T_{cp}$) and at $n=4n_0$ and $x_e=0.01$ reduces
the electrical conductivity of neutron star matter by a large factor
\begin{equation}
            {\sigma_{ep} \over \sigma_e^{\rm loc.p.}} \simeq
            {2 \cdot 10^5  \over  T^2_8}.
\label{sigmas}
\end{equation}
Such a low value of electrical conductivity would lead to
a significant decay of the electric currents, circulating
within the core with localized protons, over
a time scale of
\begin{equation}
           \tau_{\rm d}^{\rm loc.p.} \sim
           {\sigma^{\rm loc.p.}_e (R_{\rm core}^{\rm loc.p.})^2 \over c^2}
           \simeq 3 \cdot 10^7
           \left(R_{\rm core}^{\rm loc.p.} \over 10~{\rm km} \right)^2
           \,\,\, {\rm years.}
\label{taud}
\end{equation}

Current analyses of the population of radio pulsars do not
show any evidence of magnetic field decay during active lifetime
of a normal radio pulsar
\cite{BS95}. Specific {\it lower bounds} on the magnetic field
decay timescale, obtained using various types
of statistical analyses,  range from
$2\times 10^7$~years to $10^8$~years
\cite{BS95}. To be consistent with this observational fact,
we should assume that either the core with localized protons is
small, and the bulk of the field is sustained by the currents,
circulating in the outer part of the core; either the
external field is separated from the internal field, and
the observable bounds for the decay of the surface field put
no evident constraints on the evolution of the core field;
or that the core magnetic field is due not to
electric currents but results from a permanent ferromagnetic
magnetization of the matter. Actually, as demonstrated by Kutschera
and W{\'o}jcik \cite{KW89,KW92}, ferromagnetism due to a complete
spin polarization of protons and a partial spin polarization
of neutrons (of the order of $x_p$)
could be a generic property of the neutron star matter
with localized protons (for a confrontation of this theoretical prediction
with observations of radio pulsars, see \cite{HB96}).

{\it (c) Shear viscosity.}
The localization of protons
leads also to a strong decrease of
the
shear
viscosity of neutron star matter. Assuming normal neutrons, we
can estimate the value of $\eta$ of standard $npe$  matter by
the quantity $\eta^{\rm loc.p.}_{nn}$.
The localization of protons
will result in a decrease
of $\eta$ by a factor
\begin{equation}
            {\eta \over \eta^{\rm loc.p.}}
            \simeq {7 \cdot 10^5 \over T_8^2}~.
\label{etas}
\end{equation}
In contrast to the standard case
$\eta^{\rm loc.p.}$ is
temperature independent. This might be relevant for stability of
rapidly rotating neutron stars. In the standard case of the $npe$
matter, $\eta$ increases with decreasing temperature as $T^{-2}$.
Dissipative effects due to $\eta$ contribute to
damping of the secular instability driven by the
gravitational radiation reaction (GRR) \cite{F83,W84} in
rapidly rotating neutron stars. Detailed calculations show, that
viscous effects of $\eta$ damp completely the GRR secular
instability, if internal temperature falls below $10^7~$K
\cite{L95}. However, within a neutron star core with
localized protons,
the
shear viscosity remains constant and close to
the value of
the
shear viscosity of standard $npe$ matter
at $T \simeq 10^{11}~$K  [see Eq.\ (\ref{etas})].
Such a low value of
the
shear
viscosity could not prevent the growth of the GRR secular
instability at any internal temperature of a neutron star
\cite{L95}.

{\it (d) Neutrino cooling.}
At the earlier stages of the evolution
matter with localized protons cools emitting
neutrinos via two reactions involving only
nucleons, and several reactions, which involve also electrons.
The two nucleonic reactions are the neutrino--antineutrino pair
bremsstrahlung in the $nn$ and $np$ collisions.
The first process (e.g.\ \cite{FM79}) is common for both standard
and localized protons models of matter, and the rate
of energy emission in this process varies with
temperature as $T^8$. The second nucleonic
process is modified drastically by the localization
of protons: its emissivity becomes proportional
to $T^6$, reproducing the temperature dependence of
the direct URCA process. Nevertheless, the $np$
bremsstrahlung remains several orders of magnitude
less efficient than the direct URCA
($Q_{\rm Brem}^{n-{\rm loc.p.}} / Q_{\rm dURCA} \sim 3 \cdot 10^{-5}$).
On the other hand, at temperatures below $\sim 10^9$ K
its emissivity exceeds that of the modified URCA
($Q_{\rm Brem}^{n-{\rm loc.p.}} / Q_{\rm mURCA} \sim 2 \cdot 10^3~T^{-2}_8$)
process, which, in turn, is thought to govern the cooling
in the standard (non--localized) model, when
the direct URCA is forbidden by the momentum conservation
law. Thus, one might expect that cooling
of a neutron star with significant fraction of mass being
in the phase with localized protons would follow some
intermediate path between the curves
describing the standard (modified URCA) and accelerated
(direct URCA) cooling.

As is any nucleonic process,
the $np$ bremsstrahlung is subject to strong suppression,
if the neutron superfluidity appears. In this case,
the neutrino pair emission from the scattering of electrons
off localized protons serves as a dominant mechanism
of cooling of a neutron star core.

{\it (e) Final remarks.}
In this subsection we will come back to the problem of
a crystalline
ordering of localized protons and comment on the values of
the transport coefficients in this case.
As discussed in Section 2, we consider this possibility
unlikely, however, see ref.\ \cite{KW95}.
For temperatures below 1 MeV the phase space
available for phonons (in the proton crystal) is small (as $T^3$).
Then the transport of energy, charge and momentum would
be limited by $ee$, $nn$, and, in the case of
the electrical conductivity, by $en$ collisions. This means
that all the transport coefficients would be even larger
than those in the standard case and would reproduce the standard
(Fermi liquid) temperature dependences. The situation would be
complicated by the band structure of quasiparticle states,
however, this would further increase the transport coefficients
by reducing the phase space available for scattering
quasiparticles.

\vskip 0.5cm

\parindent 0pt
{\bf Acknowledgements.}
We are deeply grateful  to D.G.\ Yakovlev
for suggesting the importance of the modifications
of neutrino energy loss rates, many fruitful discussions,
and constant interest in this work.
We thank J.L.\ Zdunik for
his helpful comments concerning the role of shear viscosity in
stabilization of rapid rotation of neutron stars.
One of us (DAB)
is pleased to acknowledge excellent working conditions
at Copernicus Astronomical Center (CAMK), Warsaw, where
the major part of this work has been done. This research
has been supported in part by the KBN Grant 2 P03D 014 13,
RBRF (grant 99-02-18099), RBRF-DFG (grant 96-02-00177G), and
INTAS (grant 96-0542).

\pagebreak


\end{document}